\documentclass{svproc}
\usepackage{url}
\usepackage{graphicx}
\usepackage{color}
\begin{document}
\mainmatter              
\title{Neural Network Tire Force Modeling for Automated Drifting (AVEC '24)}
\titlerunning{Neural Network Drifting (AVEC '24)}  
%
\author{Nicholas Drake Broadbent \and
Trey Weber \and
Daiki Mori \and
J. Christian Gerdes}
\authorrunning{Nicholas Drake Broadbent et al. (AVEC '24)} 
%
\tocauthor{Nicholas Drake Broadbent, Trey Weber, Daiki Mori, and J. Christian Gerdes}
\institute{Stanford University, Stanford CA 94305, USA
\email{\{ndbroadb,tpweber,dmori,gerdes\}@stanford.edu}}

\maketitle              

\begin{abstract}
Automated drifting presents a challenge problem for vehicle control, requiring models and control algorithms that can precisely handle nonlinear, coupled tire forces at the friction limits. We present a neural network architecture for predicting front tire lateral force as a drop-in replacement for physics-based approaches. With a full-scale automated vehicle purpose-built for the drifting application, we deploy these models in a nonlinear model predictive controller tuned for tracking a reference drifting trajectory, for direct comparisons of model performance. The neural network tire model exhibits significantly improved path tracking performance over the brush tire model in cases where front-axle braking force is applied, suggesting the neural network’s ability to express previously unmodeled, latent dynamics in the drifting condition.
\keywords{Machine Learning, Tire Modeling, Autonomous Driving.}
\end{abstract}

\section{Introduction}
\vspace{-.25cm}
The maneuvering capability of a vehicle is fundamentally limited by the friction between the tires and the road. Vehicle operation at the friction limits may require large lateral and longitudinal tire slip, a regime that can be difficult to model accurately in the presence of parameter variation~\cite{Svendenius}. This is due in part to the many empirically defined characteristics of tire material composition (e.g. coefficient of friction between tire and road, cornering stiffness of the tire, thermal properties) and the geometry of tire and suspension subassemblies (e.g. camber, caster, and toe angles) that can significantly impact the overall vehicle dynamics~\cite{Pacejka}. The resulting force and moment computations of physics-based models are sensitive to the precise representation of these dynamic characteristics, particularly when operating in coupled slip regions at the limits of handling~\cite{Svendenius}.\par

Recently, autonomous racing and drifting have emerged as challenge problems for demonstrating precise vehicle control at the friction limits. While interesting problems on their own, the insights gained from automated racing and drifting also lay the foundation for future automated systems that could improve safety. To succeed, controllers must reliably control tire forces in these nonlinear, coupled slip regions. Autonomous drifting, in particular, poses challenges as the tires are not only operating in these coupled slip regions but also heating and disintegrating over the course of the test~\cite{Kobayashi}. 

Many examples in the literature have shown that an autonomous vehicle, with thoughtful modeling and control design, can drift. Velenis was one of the first to develop a controller for automated drifting, stabilizing the vehicle around an unstable cornering equilibrium with a large sideslip angle~\cite{Velenis}. Subsequent control approaches by other authors have extended this result to path tracking and, with the use of front axle braking, simultaneous velocity control, demonstrated on full scale test vehicles~\cite{Goel}. The success of these approaches, applied to a variety of autonomous drifting problems, suggest that automated vehicles could harness these dynamics for greatly increased maneuverability. 

Perhaps surprisingly, front axle braking while drifting poses an even more difficult modeling problem than the rear axle.  Unlike the rear tires, the front tires are not always saturated and the tires are not coupled through a locked differential. Front suspension geometry while drifting with large steering angles further complicates modeling the coupled front tire forces. 
 This is particularly true with dedicated drifting vehicles such as Takumi, our automated 2019 Toyota Supra built to Formula Drift specifications.  Takumi features a custom front wheel alignment designed for high-performance drifting (-7 +/- 0.3 degrees camber and 6 +/- 0.3 degrees caster at 0 degrees steering angle). This setup creates effects in the coupled slip behavior that can be difficult to model, since the tire contact patch changes size and location based on steering angle.\par

Artificial intelligence offers a chance to address some of these challenges. Djeumou \textit{et al}. developed front and rear tire force models for drifting using neural ordinary differential equations and neural-ExpTanh parameterization, ensuring physical accuracy by constraining predictions to a family of solutions and capturing higher-order effects from vehicle data. Compared to a nonlinear model predictive controller using the Fiala brush tire model, their models significantly improved tracking, smoothed control inputs, and sped up computation time in experiments~\cite{Djeumou}. Notably, their approach focused on steering and drive torque and did not  include the front axle braking necessary for independent speed control. Given the particular challenges with modeling front axle tire force generation under braking, we propose a neural network for predicting front tire lateral force that makes no prior assumptions about the shape of the resulting tire curve (or constraining predictions accordingly), relying exclusively on capturing these dynamics with vehicle data. Comparing the performance to that of the Fiala brush model in an experimental setup similar to that of Djeumou \textit{et al}, the learning-based model achieved significantly better overall trajectory tracking performance with no increase in computational complexity.  Deeper analysis of the results highlights the importance of training data coverage of the state space and potential opportunities for extending this approach to learn higher-order effects.

\section{Experimental Setup}
\vspace{-.25cm}
\subsection{Neural Network Model Development}
\begin{figure}
\vspace{-.5cm}
\includegraphics[width=0.8 \textwidth]{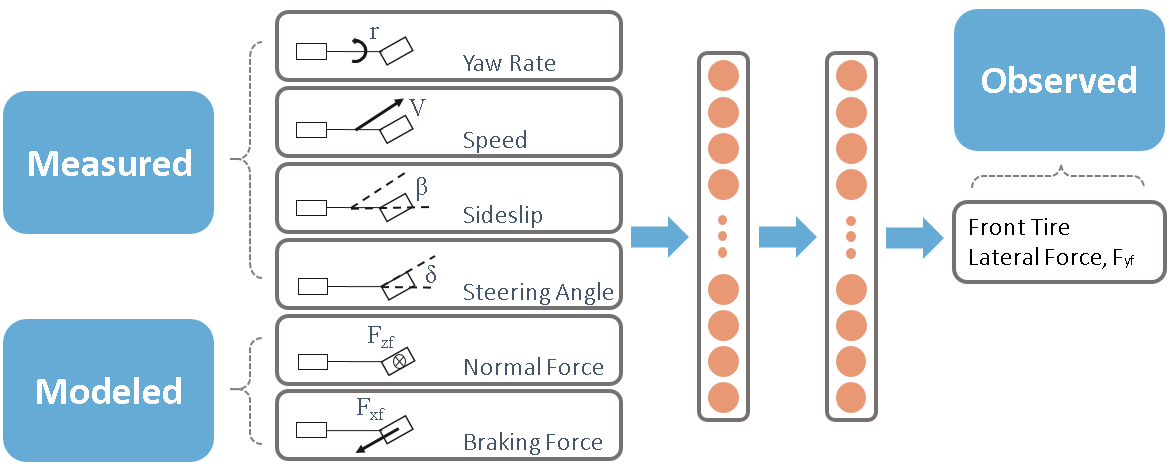}\centering
\caption{Neural network architecture for predicting lateral tire force} \label{neuralnet}
\vspace{-.5cm}
\end{figure}

\vspace{-.25cm}
We structure the input layer of the neural network around the same terms that define lateral tire force generation within the Fiala brush model, as shown in Fig. \ref{neuralnet}. We label vehicle states (yaw rate, velocity, and sideslip angle) and control inputs (steering angle and braking force) with raw measurements from the vehicle. The corresponding normal and lateral forces are labeled with estimates provided by an unknown input observer.

The data used to train the neural network features a combination of automated and manual drifting, amounting to approximately 30 minutes recorded up to one month before these comparative experiments took place. One dataset, featuring automated drifting with instances of front axle braking collected the day before these experiments, is held out of the training data in order to iteratively tune the hyperparameters of the model including batch size, training epochs, activation function, and number of hidden elements. The resulting neural network consists of a three-layer feedforward architecture with 8 elements in the first hidden layer, 16 elements in the second hidden layer and tanh activation functions in both hidden layers. 
 While quite small by neural network standards, this model size corresponds to a roughly 35ms average solve time, approximately equivalent to that of the physics-based tire model used for comparisons. Therefore, a model of this size represents a drop-in replacement for a physical tire model. Training proceeds by cycling through mini-batches of 1000 samples over 1000 epochs, with loss optimization governed by the Adam optimizer and mean squared error loss function.
\vspace{-.25cm}
\subsection{Trajectory Generation}
\vspace{-.25cm}
The same tire force observer that generates front tire lateral force labels for the neural network training assists in fitting front and rear axle tire parameters. In addition to fully defining the Fiala brush model that served as the point of comparison for these experiments, these tire parameters and the resulting model are used in computing the offline reference trajectory, similar in approach to Weber~\cite{Weber}. This trajectory features a 15 meter radius circle path with a constant sideslip angle of -40 degrees. By incorporating front axle braking, the target velocity decreases from the equilibrium value without the use of brakes ($V_{sol}$) with each revolution of the map (lap 1: $V_{des} = V_{sol}$, lap 2: $V_{des} = 0.95 \cdot V_{sol}$, lap 3: $V_{des} = 0.875 \cdot V_{sol}$), allowing us to compare model performance in the condition of increasing front axle longitudinal force (lap 1: $F_{xf,ref} = 0$ N, lap 2: $F_{xf,ref} = 1000$ N, lap 3: $F_{xf,ref} = 2150$ N).
\vspace{-.25cm}
\subsection{Control Architecture}
\vspace{-.25cm}
Nonlinear Model Predictive Control (NMPC) can handle multi-input, multi-output systems with nonlinear dynamics and constraints on both states and inputs while predicting future system behavior. These properties are advantageous in trajectory tracking for automated drifting, as exhibited by both Goel and Weber~\cite{Goel,Weber}. The implementation of NMPC for these experiments is very similar to that of the latter contribution, with a similar cost function (reformulated as a velocity tracking problem) and slightly different costs. The baseline physics-based MPC incorporates a Fiala brush front tire model. The neural network MPC (NNMPC) features an otherwise identical control framework with the same rear tire model and the learning-based front tire lateral force model as a drop-in replacement for the Fiala brush tire model.

\section{Results and Discussion}

\begin{figure}
\vspace{-.5cm}
\makebox[\textwidth][c]{\includegraphics[width=1.5\textwidth]{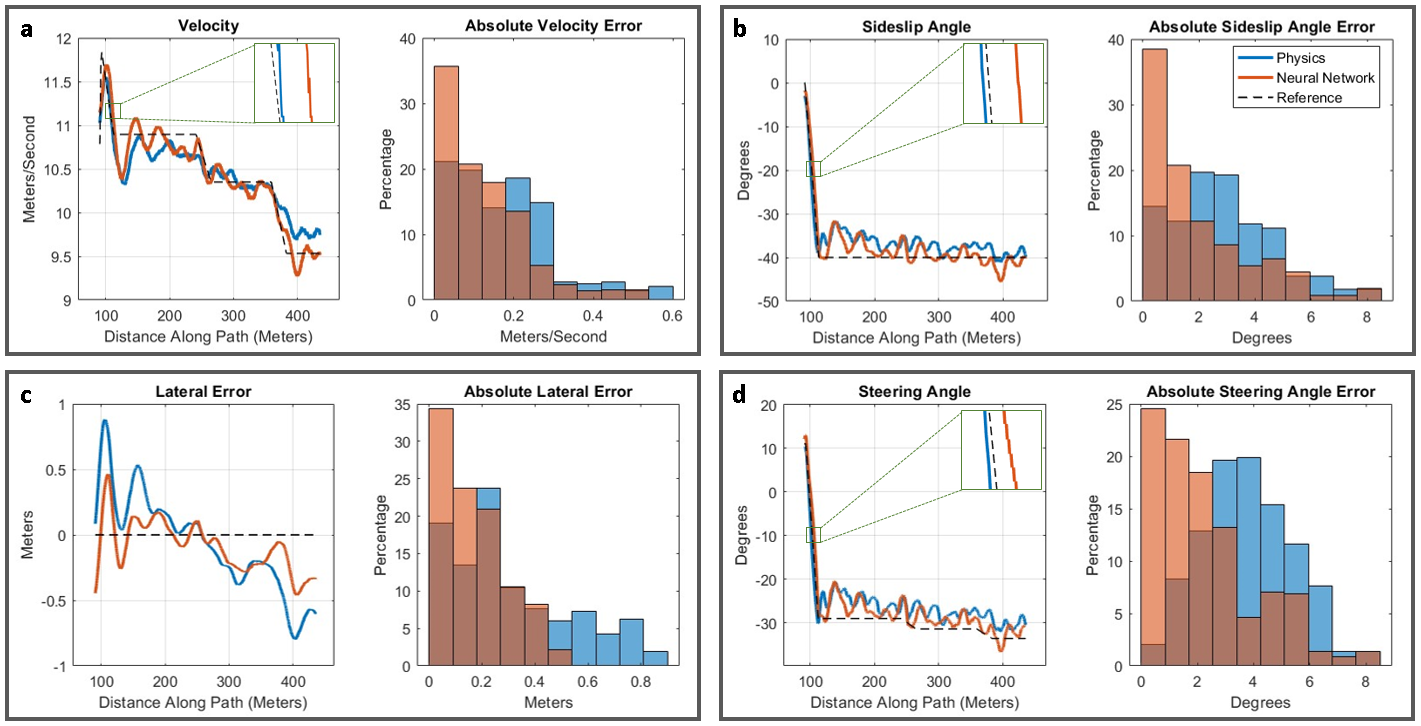}}
\caption{Tire model comparison of tracking performance under braking} \label{results}
\vspace{-.5cm}
\end{figure}

\noindent While both controllers slightly undershoot desired velocity after initiation, NNMPC is able to respond to the error more quickly and  with less oscillation, as shown in Fig. \ref{results}a. This is consistent throughout the run, whereas the physics-based MPC tends to respond to changes in desired speed more slowly, incurring a higher frequency of large absolute velocity errors in the process. This hesitation persists in sideslip angle tracking as well, where physics-based MPC shows some greater deviation from the desired -40 degree sideslip while negotiating control in the other states, as shown in Fig. \ref{results}b. Conversely, NNMPC is able to more quickly achieve and maintain the desired sideslip angle, leading to higher frequencies of small absolute sideslip angle errors in the process.\par

NNMPC’s trend of high performance in the velocity states translates well to path tracking performance, where it exhibits a relatively low mean and max absolute lateral error, as shown in Fig. \ref{results}c. In contrast, the physics-based control appears to cause Takumi to slowly slide out from the desired path as the experiment progresses. This trend is consistent with the fact the tire temperature increases throughout the experiment and proportionally reduces friction, as shown by Kobayashi~\cite{Kobayashi}. Conversely, the neural network-based model does not rely upon explicit tire parameterization for the front axle. The neural network may potentially be underfitting these temperature-dependent friction dynamics by generalizing to tire force generation characteristics that are indicative of a wide range of tire temperatures. \par

NNMPC’s comparatively strong performance trends in both velocity and path state tracking appears to yield an overall reduced steering control effort required to maintain the drift equilibrium throughout the maneuver, as shown in Fig. \ref{results}d. However, if we decompose the stages of this experiment further into the drift initiation region (s = 90.7:112.5) and steady state equilibrium region (s = 112.5:435.3), we gain further insights into the advantages and disadvantages of each respective modeling approach—particularly when we focus into the initiation region dynamics, as shown in the insets of Fig. \ref{results}. For example, while it may appear that NNMPC is outperforming physics-based MPC in the initiation region, the mean absolute errors of velocity, sideslip angle, and steering angle are higher with NNMPC than with physics-based MPC—and the percent difference is significant (49\%, 31\%, and 26\%, respectively). This is in stark contrast to the steady state equilibrium region of the experiment, where the mean absolute errors of velocity, sideslip angle, lateral error, and steering angle are lower with NNMPC than with physics-based MPC, where the percent difference is significant once again (41\%, 46\%, 55\%, and 53\%, respectively). One explanation for this behavior may be found in the way in which the neural network was trained. Of the approximately 30 minutes of data used to train the model, less than 5\% can be prescribed to the drift initiation region. This imbalance in data representation can potentially lead to a bias toward solely capturing the dynamics of the steady state equilibrium region. Since the gradients calculated from the dominant region will have a greater influence on the network's parameter updates, this can cause the model to prioritize minimizing the loss in the steady state equilibrium region at the expense of capturing the dynamics of the drift initiation region, further exacerbating the imbalance represented in the data.\par

Another explanation for this behavior may be rooted in how the features and targets were labelled and synchronized. Latencies inherent in observers such as the one used to label tire force targets can cause a temporal misalignment between the observed states and the actual system states—impairing the neural network's ability to learn the correct temporal patterns and dynamics of the system. This may be particularly crucial in the drift initiation region, where the vehicle is highly dynamic, undergoing comparatively far greater velocity state derivatives (in yaw rate, velocity, and sideslip angle) than those indicative of the steady state equilibrium region.

\section{Conclusion}
\vspace{-.25cm}
This investigation presents a novel neural network architecture for predicting front tire lateral forces as a substitute for traditional physics-based models, with a specific focus on autonomous vehicle drifting maneuvers. Through comparative experimentation using a full-scale automated vehicle, we demonstrated that the neural network model significantly enhances path tracking performance, especially under conditions involving front-axle braking forces. The implications of this study are significant for the development of advanced control systems in autonomous vehicles, particularly those designed to operate in extreme conditions. As we continue to build trust and understanding in machine learning techniques, we may be able to achieve higher levels of precision and reliability in vehicle dynamics modeling, paving the way for safer and more efficient autonomous driving technologies.\par

This research may be extended in several ways to ultimately achieve similar closed-loop performance in trajectories of increasing complexity. Since observer latency and temporal misalignment of labeled data may have been an issue with this approach, we are currently investigating approaches with target labeling that rely solely upon vehicle-collected measurements to potentially eliminate this behavior. Additional performance enhancements can conceivably be obtained with the inclusion of additional relevant states as input to the neural network (e.g. tire temperature, in order to capture temperature-dependent dynamics) or simply expanding the complexity of the network itself as computational limitations allow.

%
%
%
%

\end{document}